\documentclass[10pt,notitlepage]{article}
\usepackage{amsmath}
\usepackage{amssymb}
\usepackage{graphics}
\usepackage{graphicx}

\def\func#1{\mathop{\rm #1}\nolimits}

\def\ii{\'{\i}}

\def\ca{\c{c}\~{a}}

\def\ro{\hat{\rho}}
\def\tr#1{\func{tr}#1}
\def\d{\mathrm{d}}
\def\e{\mathrm{e}}

\def\adag{\hat{a}^{\dagger}}
\def\a{\hat{a}}

\def\ro{\hat{\rho}}
\def\rot#1{\hat{\rho}\left(#1\right)}
\def\delA{\partial_{\alpha}}
\def\delAcc{\partial_{\alpha^*}}
\def\delB{\partial_{\beta}}
\def\delBcc{\partial_{\beta^*}}
\def\delt{\partial_t}
\def\so#1{\mathcal{#1}}
\def\cP#1#2{\left[#1,#2\right]_P}

\begin{document}

  \title{Time evolution of the classical and quantum mechanical versions 
    of diffusive anharmonic oscillator: an example of Lie algebraic techniques}
  \author{
  	J. G. Peixoto de Faria\thanks{Electronic address: jgpfaria@des.cefetmg.br}\\
  	\footnotesize{Departamento Acad\^{e}mico de Disciplinas B\'{a}sicas,}\\
  	\footnotesize{Centro Federal de Educa{\ca}o Tecnol\'{o}gica de Minas Gerais,}\\
  	\footnotesize{Belo Horizonte, MG, 30510-000, Brazil}
  	} 
  
\maketitle  
  
  \begin{abstract}
    We present the general solutions for the classical and quantum
    dynamics of the anharmonic oscillator coupled to a purely 
    diffusive environment. In both cases, these solutions are obtained by
    the application of the Baker-Campbell-Hausdorff (BCH) formulas to
    expand the evolution operator in an ordered product of 
    exponentials. Moreover, we obtain an expression for the Wigner
    function in the quantum version of the problem. We observe that
    the role played by diffusion is to reduce or to attenuate the 
    the characteristic quantum effects yielded by the nonlinearity,
    as the appearance of coherent superpositions of quantum states
    (Schr\"{o}dinger cat states) and revivals. 
  \end{abstract}
% 
%
%  \keywords{
%    Anharmonic oscillator \and Quantum systems \and Decoherence \and Lie algebras.}
%  \PACS{{03.65.Yz}{Decoherence; open systems; quantum statistical methods} \and
%  			{02.20.Sv}{Lie algebras of Lie groups}}
%	}
%	\authorrunning{Peixoto de Faria}
%	\titlerunning{Time evolution of the classical and quantum...}

\section{Introduction}
In the last decades, the investigation about the transition from quantum
to classical dynamics has progressed enormously. This was induced in part
by the development of the experimental techniques, especially in quantum
optics \cite{paris}, and in part by the possibility of appearance of 
technology in quantum information processing \cite{nielsen00}, \cite{gisin02}. 
Such research aims to understand how the typically 
quantum effects disappear in the dynamics of macroscopic systems. 
According to a popular theoretical model, one believes that 
the emergence of the classical world from quantum mechanics is
a consequence of the unavoidable coupling between the macroscopic system
and its environment. For example, in accordance with this proposal \cite{decoherence}, environmental
coupling is responsible by the rapid evolution of coherent quantum superpositions
of macroscopically distinguishable states (Schr\"odinger cat states) into
statiscal mixtures, a phenomenon known as decoherence.

Despite these theoretical advances, some aspects of the quantum to
classical transition remain subtle and controversial, especially
in the case of classically nonlinear or chaotic systems. For some
authors, the departure of the quantum mean values of observables from
the corresponding classical ones (correspondence breakdown) in chaotic
systems occurs in a very small time scale (see, for example, \cite{habib98}) 
and they sustain 
the idea that the coupling with purely diffusive environment can reduce
these discrepancies and increase the break time by decoherence. On the
other hand, for other authors (see \cite{ball05}) decoherence is not
necessary to explain the classical behavior of macroscopic systems 
(including the chaotic ones) since the observed discrepancies between
quantum and classical mean values of observables are negligible for
any current realistic measurement.

In order to shed some light on this debate, we turn our attention to the
one-dimensional nonlinear or anharmonic oscillator (AHO) 
coupled to a purely diffusive reservoir.
Three are the reasons for the choice of this model. Firstly, in the limit of 
vanishing environmental coupling, its quantum dynamical evolution exhibits 
several effects without analogous in its classical counterpart, such as 
revivals and appearance of coherent superpositions of states \cite{yurke86}.
Secondly, by virtue of its relative simplicity, it is possible to obtain
the exact solutions of the equations of motion for quantum 
\cite{milburn86,milburn_holmes,daniel_milburn,perinova90,chaturvedi91,kheruntsyan99,
berman04} and classical versions
of the model even in the presence of the reservoir. Last, but not least, the recent
technical advances in trapping and controlling cold atoms suggest 
the Bose-Einstein condensates (BEC) as potential candidates to implement 
experimental tests of this model. In fact, in the single-mode 
approximation, a BEC trapped in a optical lattice
is suitably described by the quantum nonlinear oscillator \cite{BEC}.
Hence, dissipative AHO became a largely studied model in the literature.
In Ref. \cite{daniel_milburn}, Daniel and Milburn obtained the exact
evolution of the \textit{Q} function associated to an initial 
coherent state of an AHO subject to attenuation or amplification.
The authors showed that the effects yielded by nonlinearity,
such as revivals and squeezing, gradually vanish if a non-unitary
mechanism is taken into account. These results were extended by
Pe$\check{\mathrm{r}}$inov\'{a} and Luk$\check{\mathrm{s}}$ \cite{perinova90} to
an arbitrary initial state. In Ref. \cite{kheruntsyan99}, Kheruntsyan obtained
the steady state of Wigner function of a single driven damped cavity mode
in the presence of a Kerr medium. The author studied a model of 
reservoir that included two-photon absortion, besides the usual one-photon
absorption.
Closely related to the present contribution, Chaturvedi and Srinivasan
\cite{chaturvedi91} found an exact solution of a class of master equations
governing the dynamics of a chain of coupled dissipative AHO. The authors
used thermofield dynamics notation in order to map a master equation into
a Schr\"{o}dinger equation with a non-hermitean hamiltonian. The form of 
the general solution of this equation was expressed as an ordered 
product of exponentials of operators acting on an arbitrary initial 
condition.

In this contribution, we obtain the exact time evolutions of the density
operator (section 2) and of the classical distribution function (section 3)
in the quantum and classical versions of the nonlinear oscillator coupled to 
a purely diffusive environment. Both results are obtained by the
application of the Lie algebraic techniques, in particular, the 
Baker-Campbell-Hausdorff (BCH) formulas used to expand a Lie exponential in
an ordered product of exponentials \cite{chaturvedi91,kumar65,wilcox67,gilmore74,witschel81}. 
In the quantum case, the result allows us to 
find the corresponding Wigner function in terms of an expansion in associated
Laguerre polynomials. It is important to mention that the Wigner function
should smoothly approach the corresponding classical distribution in the 
appropriate limit \cite{habib98}. Thus, we take the classical limit of the partial differential
equation (PDE) that governs the time evolution of the Wigner function and we 
obtain a classical Fokker-Planck (FP) equation. The algebraic structure of 
the master equation in the quantum case is preserved by the corresponding
FP equation in the classical case, so that we can extend the methods 
applied to obtain the solution of the first in the finding of the 
solution of the last. 

In Ref. \cite{oliveira06}, Oliveira and co-workers investigated the
diffusive AHO and showed that the 
break times (\textit{i.e.}, the characteristic times of departure 
of the quantum and classical dynamics) depend strongly on observable and
initial condition. A more ``fair" comparison between the two dynamics
shall be given by the evaluation of the distance between the corresponding distributions
in the phase space \cite{toscano05}. 
Hence, the exact solutions of the equations of motion for the Wigner function
and classical distribution function will allow to define analytically the
break time for the diffusive AHO and its dependence in terms of the
nonlinearity strength and diffusion constant. 

We conclude this work by comparing the quantum and classical evolutions of the
Wigner and the classical distribution functions for an initial
coherent state in the AHO with and without diffusion (section 4). The results
are preliminar and deserve a more careful analysis, but they suggest that, as
expected, in quantum case, inclusion of diffusion reduces the phase space interferences 
and therefore prevent the
appearance of quantum coherent superpositions. The area of the regions
where the Wigner function should be negative is reduced too. As time goes by,
the Wigner distribution gradually takes the form of an annular volume around
the origin of the phase space. For later times, this volume becomes more ``fat''
and ``flat''. In the classical case, the fine-structured whorl yielded by 
the distribution in the diffusionless regime is destroyed. As it
happens in the quantum case, diffusion turns the classical distribution more
``fat'' and ``flat'' around the origin. These results suggest that
the Wigner function of the
quantum diffusive AHO converges gradually to the
distribution function of the corresponding classical version of the model
in non-unitary evolution. 

\section{Quantum mechanical diffusive anharmonic oscillator}

Let us consider the AHO coupled to a thermal bath of oscillators
in equilibrium at temperature $T$. Assuming that the nonlinearity strength is small and
the coupling to the reservoir degrees of freedom is weak, we obtain the
following master equation in the interaction picture:
\begin{eqnarray}
  \overset{.}{\ro }\left( t\right) &=&\mathcal{L}\rho\left( t \right) 
  \label{master1}  \\
  &=& -ig\left[\left(\adag\a\right)^2,\rot{t}\right]+
  k\left(\bar{n} + 1\right)\left[2\a \rot{t}\adag - \adag\a \rot{t}
  - \rot{t}\adag\a\right] \notag \\
  && + k\bar{n}\left[2\adag \rot{t}\a - \a\adag \rot{t}
  - \rot{t}\a\adag\right] \notag \, ,
\end{eqnarray}
where $k$ and $g$ are the damping and nonlinearity constants, respectively, 
and $\bar{n}$ is the average number
of thermal photons in the mode $\omega$ of reservoir ($\omega$ is the
natural frequency of the oscillator). 
The density operator $\ro \left( t\right)$ represents the state of the
system at time $t$; $\a$ and $\adag$ are the annihilation and creation 
operators, respectively.
We are interested in the so called diffusive limit of the above equation. 
This limit is obtained by taking the damping constant going to zero,
$k\rightarrow 0$, and the number of thermal photons going to infinite,
$\bar{n} \rightarrow \infty$, keeping the product $\kappa = k\bar{n}$
finite. 
Thus, the master equation (\ref{master1}) becomes
\begin{eqnarray}
\overset{.}{\ro }\left( t\right) &=&\mathcal{L_\infty}\rho\left( t \right) 
  \label{master2}  \\
  &=& -ig\left[\left(\adag\a\right)^2,\rot{t}\right] \notag \\
  && + 2\kappa\left[\a \rot{t}\adag + \adag \rot{t}\a - \adag\a \rot{t}
  - \rot{t}\adag\a- \rot{t}\right] \notag \\
  &=& -ig\left( \mathcal{M}^{2}-\mathcal{P}^{2}\right) 
  \rot{t}  +2\kappa \left[ \mathcal{J}+\mathcal{R}-\left( \mathcal{M}+\mathcal{P}%
  +1\right) \right] \rot{t}   \notag \, .
\end{eqnarray}

Using the notation given in Ref. \cite{wang02},
the super-operators (sup-op) that appear in the last line of above equation are
defined by 
\begin{subequations}
\label{supop}
  \begin{align}
    \mathcal{M}&\equiv \left( \adag\a\right)_{l}=
    \adag_{l}\a_{l}\, , \label{supopM} \\ 
    \mathcal{P}&\equiv \left( \adag\a\right)_{r}=
    \a_{r}\adag_{r}\, , \label{supopP} \\
    \mathcal{J}&\equiv \a_{l}\adag_{r}=\adag_{r}\a_{l} \, ,
    \label{supopJ} \\
    \mathcal{R}&\equiv \a_{r}\adag_{l}=\adag_{l}\a_{r}
    \label{supopR} \, . 
  \end{align}
\end{subequations}
Here,  $\a_{l}$, $\a_{l}^{\dagger }$, $\a_{r}$, $%
\a_{r}^{\dagger }$ represent the left and right actions of the creation and
annihilation operators on a generic operator $\hat{O}$:
\begin{equation}
  \a_{l}\hat{O} = \a\hat{O}\, ,\: \adag_{l}\hat{O} = \adag\hat{O}\, ,\:
  \a_{r}\hat{O} = \hat{O}\a \,, \:\adag_{r}\hat{O} = \hat{O}\adag \, .
  \label{RandLsupop}
\end{equation}

At this point, it is convenient to introduce the following nomenclature.
We call $\mathcal{B}\left(\mathcal{H}\right)$ the set of operators 
that act on the oscillator
space state $\mathcal{H}$. The elements of  $\mathcal{B}\left(\mathcal{H}\right)$ 
can be assigned to vectors of an extended Hilbert space constructed
by direct product between the original space state and its dual
$\mathcal{H}^*$, \textit{viz.} $\mathcal{H} \otimes
\mathcal{H}^*$. This extended Hilbert space is frequently called 
Hilbert-Schmidt space or Liouville space \cite{royer}.

The formal solution of Eq. (\ref{master2}) is given by
\begin{equation}
  \rot{t}  = \e^{\mathcal{L}_{\infty }t}\rho \left( 0\right) \, ,
  \label{formal1}
\end{equation}
where $\rot{0}$ represents the initial state of the AHO. 
The evolution of a generic initial state 
\begin{equation}
  \rot{0} = \sum_{m,n} \rho_{m,n} \left|m\right\rangle\left\langle n \right| 
  \label{initial-state}
\end{equation}
can be evaluated by expanding the exponential $\e^{\mathcal{L}_\infty t}$
in an ordered product of exponentials. Usually, this task is achieved by 
the systematical application of the Lie algebraic methods, in 
particular, using the
Baker-Campbell-Hausdorff (BCH) formulas 
\cite{kumar65,wilcox67,gilmore74,witschel81,steinberg}. For the
dissipative AHO, this expansion was obtained in 
Ref. \cite{chaturvedi91}, and we reproduce in detail the procedure here. 
To carry out this expansion, we begin by evaluating the commutation relations
between the sup-op defined in (\ref{supop}). They are listed in Table \ref{comutsupop}.
\begin{table}[tbh]
  \centering
    \begin{tabular}{c|c c c c}
      &$\so{M}-\so{P}$ &$\frac{1}{2}\left(\so{M}+\so{P}+1\right)$ &$\so{R}$ &$\so{J}$  
      \\ \hline
      $\so{M}-\so{P}$                            &0 &0        &0                &0                              \\ 
      $\frac{1}{2}\left(\so{M}+\so{P}+1\right)$  &0 &0        &$\so{R}$         &$-\so{J}$                      \\ 
      $\so{R}$                                   &0 &$-\so{R}$&0                &$-\left(\so{M}+\so{P}+1\right)$\\
      $\so{J}$                                   &0 &$\so{J}$ &$\so{M}+\so{P}+1$&0      
    \end{tabular}
  \caption{Commutation relations between the sup-op defined in Eq. (\ref{supop}).
    The $i,j$ entry in the table is the result of the commutation between the
    sup-op of the $i$-th row and the sup-op of the $j$-th column.}
  \label{comutsupop}
\end{table}

The sup-op $\mathcal{M}-\mathcal{P}$, $\frac{1}{2}\left( \mathcal{M}+\mathcal{P}+1\right) $, 
$\mathcal{J}$, $\mathcal{R}$ form a four-dimensional Lie algebra, which 
we will denominate $\mathcal{A}_4$. We easily recognize a subalgebra
$su\left(1,1\right)$ contained in $\mathcal{A}_4$,
formed by the set $\frac{1}{2}\left( \mathcal{M}+\mathcal{P}+1\right) $, 
$\mathcal{J}$, $\mathcal{R}$. Let us rewritten the Liouvillian $\so{L}_{\infty }$ 
as 
\begin{equation}
  \so{L}_{\infty }=ig\left( \so{M}-\so{P}\right) -
  \Lambda\left( \mathcal{M}+\mathcal{P}+1\right)
  +2k\left( \mathcal{J}+\mathcal{R}\right) \, ,
  \label{liouvq}
\end{equation}
where we define the sup-op
\begin{equation}
  \Lambda= ig\left( \mathcal{M}-\mathcal{P}\right) +2\kappa \, ,
  \label{Lambda1}
\end{equation}
that is to be formally considered a c-number, since the sup-op $\so{M}-\so{P}$
commutates with the rest. The eigenvectors of $\Lambda $
are the same of $\so{M}-\so{P}$. It is easy to verify that these 
eigenvectors belong to the set 
$\left\{ \left| m\right\rangle \left\langle n\right| \right\} $, with
eigenvalues $ig\left( m-n\right) +2\kappa $.

The formal solution (\ref{formal1}) can be written as
\begin{equation}
\ro \left( t\right) =\exp \left[ -\Lambda t\left( \mathcal{M}+%
\mathcal{P}+1\right) +2\kappa t\left( \mathcal{J}+\mathcal{R}\right) \right]
\exp \left[ igt\left( \mathcal{M}-\mathcal{P}\right) %
\right] \ro \left( 0\right) \, .
\label{formal2}
\end{equation}
Applying the well-known BCH formulas to expand $su\left(1,1\right)$
Lie exponentials \cite{witschel81,wod85,dattoli88}, we rewritten the above expression as
\begin{eqnarray}
  \ro \left( t\right) &=&\exp \left[ \Gamma \left( t\right) \mathcal{R}\right]
  \exp \left\{ \ln \left[ \Gamma _{0}\left( t\right) \right] \left( \mathcal{M}%
  +\mathcal{P}+1\right) \right\} \exp \left[ \Gamma \left( t\right) \mathcal{J}%
  \right] \notag \\
  &&\times \exp \left[ igt\left( \mathcal{M}-\mathcal{P}\right) 
  \right] \ro \left( 0\right) \, , \label{formal3}
\end{eqnarray}
where $\Gamma \left( t\right) $, and $\Gamma _{0}\left( t\right) $ are
time-dependent sup-op given by 
\begin{subequations}
  \label{Gammas}
  \begin{align}
    \Gamma _{0}\left( t\right) =\frac{\Delta}{\Delta \cosh \left( \Delta
    t\right) +\Lambda \sinh \left( \Delta t\right) }\, , \label{Gamma0} \\
    \Gamma\left( t\right) =\frac{2\kappa \sinh \left( \Delta t\right) }
    {\Delta \cosh\left( \Delta t\right) +\Lambda \sinh 
    \left( \Delta t\right) } \, . \label{Gamma_}
  \end{align}
\end{subequations}
Here, we define 
\[
  \Delta =\sqrt{\Lambda ^{2}-4\kappa ^{2}}.
\]

Let us consider the initial state (\ref{initial-state}). Its time evolution
is
\begin{eqnarray*}
  \rot{t} &=&\exp \left[ \Gamma \left( t\right) \mathcal{R}\right]
  \exp \left\{ \ln \left[ \Gamma _{0}\left( t\right) \right] \left( \mathcal{M}%
  +\mathcal{P}+1\right) \right\} \exp \left[ \Gamma \left( t\right) \mathcal{J}%
  \right] \\
  &&\times \sum_{m,n}\rho _{mn}\exp \left[ igt\left( m-n\right) 
  \right] \left| m\right\rangle \left\langle n\right| .
\end{eqnarray*}
The action of $\exp \left[ \Gamma \left( t\right) \mathcal{J}\right] $
on $\left| m\right\rangle \left\langle n\right| $ is obtained in the 
following way:
\begin{eqnarray}
  \sum_{m,n}\exp \left[ \Gamma \left( t\right) \mathcal{J}\right] \left|
  m\right\rangle \left\langle n\right| &=&\sum_{j}\sum_{m,n}\frac{\Gamma
  ^{j}\left( t\right) }{j!}a^{j}\left| m\right\rangle \left\langle n\right|
  \left( a^{\dagger }\right) ^{j}  \label{aux1} \\
  &=&\sum_{m,n}\sum_{j=0 }^{\left( m,n\right) }\sqrt{\frac{m!n!}{\left(
  m-j\right) !\left( n-j\right) !}}\frac{\Gamma ^{j}\left( t\right) }{j!}%
  \left| m-j\right\rangle \left\langle n-j\right|  \notag \\
  &=&\sum_{m,n}\sum_{j=0 }^{\left( m,n\right) }\sqrt{\frac{m!n!}{\left(
  m-j\right) !\left( n-j\right) !}}\frac{\gamma_{m-n} ^{j}\left(t\right) }{j!}%
  \left| m-j\right\rangle \left\langle n-j\right| ,  \notag
\end{eqnarray}
where $(m,n)=\min\left(m,n\right)$, and $\gamma_n \left( t\right) $ is a 
c-number function obtained by
direct substitution of the sup-op $\Lambda $ by $ign+2\kappa $ in the 
expression (\ref{Gamma_}). Let us redefine the indexes
$m\rightarrow m-j$, $n\rightarrow n-j$. Thus, 
\begin{eqnarray}
  \rot{t} &=&\exp \left[ \Gamma \left( t\right) \mathcal{R}\right]
  \exp \left\{ \ln \left[ \Gamma _{0}\left( t\right) \right] \left( \mathcal{M}%
  +\mathcal{P}+1\right) \right\}  \label{aux2} \\
  &&\times \sum_{j}\sum_{m,n}\rho _{m+j,n+j}\sqrt{\frac{\left( m+j\right)
  !\left( n+j\right) !}{m!n!}}\exp \left[ igt\left( m-n\right) \right]   \notag \\
  &&\times \frac{\gamma ^{j}_{m-n}\left( t\right) }{j!}\left| m\right\rangle
  \left\langle n\right| .  \notag
\end{eqnarray}

The action of $\exp \left\{ \ln \left[ \Gamma _{0}\left( t\right) \right]
\left( \mathcal{M}+\mathcal{P}+1\right) \right\} $ on 
$\left| m\right\rangle \left\langle n\right| $ is evaluated in analogue way: 
\begin{equation*}
  \exp \left\{ \ln \left[ \Gamma _{0}\left( t\right) \right] \left( \mathcal{M}%
  +\mathcal{P}+1\right) \right\} \left| m\right\rangle \left\langle n\right|
  =\Gamma _{0}\left( t\right) \sum_{j}\frac{\left\{ \ln \left[ \Gamma
  _{0}\left( t\right) \right] \right\} ^{j}}{j!}\left( \mathcal{M}+\mathcal{P}%
  \right) ^{j}\left| m\right\rangle \left\langle n\right| .
\end{equation*}
$\left| m\right\rangle \left\langle n\right| $ is 
an eigenstate of $\mathcal{M}+\mathcal{P}$ with eigenvalue $\left( m+n\right) $.
Hence, 
\begin{eqnarray*}
  \exp \left\{ \ln \left[ \Gamma _{0}\left( t\right) \right] \left( \mathcal{M}%
  +\mathcal{P}+1\right) \right\} \left| m\right\rangle \left\langle n\right|
  &=&\Gamma _{0}\left( t\right) \sum_{j}\frac{\left\{ \ln \left[ \Gamma
  _{0}\left( t\right) \right] \right\} ^{j}}{j!}\left( m+n\right) ^{j}\left|
  m\right\rangle \left\langle n\right| \\
  &=&\Gamma _{0}^{m+n+1}\left( t\right) \left| m\right\rangle \left\langle
  n\right| =\zeta_{m-n} ^{m+n+1}\left( t\right) \left| m\right\rangle
  \left\langle n\right| \, .
\end{eqnarray*}
Here, $\zeta_n \left( t\right) $ is a c-number function obtained by the direct 
substitution of the sup-op\ $\Lambda $ by $ign+2\kappa $ 
in the expression (\ref{Gamma0}).
Substituting this result on Eq. (\ref{aux2}), we have 
\begin{eqnarray}
  \rot{t} &=&\exp \left[ \Gamma \left( t\right) \mathcal{R}\right]
  \notag \\
  &&\times \sum_{j}\sum_{m,n}\rho _{m+j,n+j}\sqrt{\frac{\left( m+j\right)
  !\left( n+j\right) !}{m!n!}}\exp \left[ igt\left( m-n\right) \right]  
  \label{aux3} \\
  &&\times \zeta_{m-n} ^{m+n+1}\left( t\right) \frac{\gamma_{m-n} ^{j}
  \left(t\right) }{j!}\left| m\right\rangle \left\langle n\right| .  \notag
\end{eqnarray}

The action of $\exp \left[ \Gamma \left( t\right) \mathcal{R}\right] $
on $\left| m\right\rangle \left\langle n\right| $ is evaluated in the following way:
\begin{eqnarray*}
  \exp \left[ \Gamma \left( t\right) \mathcal{R}\right] \left| m\right\rangle
  \left\langle n\right| &=&\sum_{l}\frac{\Gamma ^{l}\left( t\right) }{l!}%
  \left( a^{\dagger }\right) ^{l}\left| m\right\rangle \left\langle n\right|
  a^{l} \\
  &=&\sum_{l}\frac{\gamma ^{l}_{m-n}\left( t\right) }{l!}\sqrt{\frac{\left(
  m+l\right) !\left( n+l\right) !}{m!n!}}\left| m+l\right\rangle \left\langle
  n+l\right| .
\end{eqnarray*}
Substituting this result in Eq. (\ref{aux3}), we finally have the general 
solution of Eq. (\ref{master2}) with the initial condition (\ref{initial-state}):
\begin{eqnarray}
  \rot{t}  &=&\sum_{l}\sum_{j}\sum_{m,n}\rho _{m+j,n+j}\frac{\sqrt{%
  \left( m+j\right) !\left( n+j\right) !\left( m+l\right) !\left( n+l\right) !}%
  }{m!n!l!j!}  \notag \\
  &&\times \gamma_{m-n} ^{l+j}\left( t\right) \zeta_{m-n} ^{m+n+1}\left( t\right)
  \exp \left[ igt\left( m-n\right) \right]
  \left| m+l\right\rangle \left\langle n+l\right|\,  .  \label{solution1} 
\end{eqnarray}

\subsection{The Wigner function of the diffusive anharmonic oscillator}

We can obtain a representation of operators that belong to  
$\mathcal{B}\left(\mathcal{H}\right) $ as functions of the set
$\mathcal{F}\left(\Omega\right) $, i.e., functions on the 
phase space $\Omega$. This representation can be 
seen as an invertible mapping between 
$\mathcal{B}\left(\mathcal{H}\right) $ and the set
$\mathcal{F}\left(\Omega\right) $. 
One of these representations is given by the 
Weyl-Wigner transform \cite{wigner32,wigner84,lee}, defined on a generic 
operator $\hat{O} $ as
\begin{equation}
  O\left(q, p\right)=\int_{-\infty}^{\infty}
  e^{i u p/\hbar} \left\langle q - u/2\right|\hat{O}
  \left|q + u/2\right\rangle \d u  \equiv \mathcal{W}
  \left(\hat{O}\right) \left(q,p\right) \: ,
  \label{weylwigner}
\end{equation}
where $q$ and $p$ are phase space coordinates position and momentum. 
The Weyl-Wigner transform of the operator density, defined as
\begin{equation}
  W\left(q,p\right) = \frac{1}{2\pi \hbar}\int_{-\infty}^{\infty}
  e^{i u p/\hbar} \left\langle q - u/2\right|\ro
  \left|q + u/2\right\rangle \d u  \equiv \mathcal{W}
  \left(\frac{\ro}{2\pi\hbar} \right) \left(q,p\right) \: ,
  \label{wignerfunc1}
\end{equation}
yields a quasi-probability distribution function -- the Wigner function, $W\left(q, p\right)$. 
We can merge the phase space coordinates $q$ and $p$ in an unique 
complex variable $\alpha$ as follows
\[
  \alpha = q\sqrt{\frac{M\omega }{2\hbar }}+ip\sqrt{\frac{1}{2M\omega \hbar }}
  \equiv \frac{Q + iP}{\sqrt{2}} \, .
\]
Here, $M$ is the mass of the oscillator, and $Q$ and $P$ are adimensional real variables.
Adopting this representation, the Wigner function is defined by
\begin{equation}
  W\left(\alpha\right) = \frac{1}{\pi\hbar}\tr \left[\ro \hat{D}\left(\alpha\right)\e^{i\pi \adag\a}
  \hat{D}^{\dagger}\left(\alpha\right) \right]\, ,
  \label{wignerfunc2}
\end{equation}
where $\hat{D}\left(\alpha\right) = \exp\left(\alpha\adag - \alpha^* \a\right)$ is
the unitary displacement operator \cite{perelomov}, and $\tr\left(\cdot\right)$
stands for trace. 

The application of the Weyl-Wigner transform in expression (\ref{solution1}) produces
\begin{eqnarray}
  W\left(t\right)&=&\sum_{l}\sum_{j}\sum_{m,n}\rho _{m+j,n+j}\frac{\sqrt{%
  \left( m+j\right) !\left( n+j\right) !\left( m+l\right) !\left( n+l\right) !}%
  }{m!n!l!j!}  \notag \\
  &&\times \gamma_{m-n} ^{l+j}\left( t\right) \zeta_{m-n}^{m+n+1}\left( t\right)
  \exp \left[ igt\left( m-n\right) \right]\Pi_{m+l,n+l}, \label{wigner1} 
\end{eqnarray}
Hence, the Wigner function $W\left( t\right)$ of the nonlinear oscillator
is expressed in terms of the functions $\Pi_{m,n}$, that are obtained
by the Weyl-Wigner transform of the eigenfunctions $\left\{\left|m\right\rangle 
\left\langle n\right| \right\}_{m,n =0,1,\ldots }$ of the sup-op
$\left(\mathcal{M} - \mathcal{P}\right) $, i.e.
\begin{equation}
 \Pi_{m,n}\left(\alpha\right) = \left(2\pi\hbar\right)^{-1}\mathcal{W}\left(\left| m\right\rangle 
 \left\langle n\right|\right)\left(\alpha\right) \label{Pi_mn1} 
 = \frac{\left(-1\right)^m}{ \pi \hbar}\left\langle n \right| \hat{D}\left(2\alpha\right)
 \left|m \right\rangle  .
 \notag
\end{equation}
The matrix elements of the operator $\hat{D}\left(2\alpha\right)$ are
given by (see appendix B of Ref. \cite{perelomov})
\begin{equation}
	\left\langle n \right| \hat{D}\left(2\alpha\right)
	\left|m \right\rangle =
	\begin{cases}
		\sqrt{\frac{m!}{n!}}\e^{-2\left|\alpha\right|^2}\left(2\alpha\right)^{n-m} 
		L_m^{n-m}\left(4\left|\alpha\right|^2\right),\: n\geq m \\
		\sqrt{\frac{n!}{m!}}\e^{-2\left|\alpha\right|^2}\left(-2\alpha^*\right)^{n-m} 
		L_n^{m-n}\left(4\left|\alpha\right|^2\right),\: m> n 
	\end{cases}
	\label{Dnm}
\end{equation}
where $L_m^{n-m} \left(x\right)$ is the associated Laguerre polynomial.
Hence, we have
\begin{subequations}
 \label{Pi_mn2}
 \begin{align}
  \Pi_{m,n}\left(\alpha\right) &=\frac{\left(-1\right)^m}{\pi\hbar}
  \sqrt{\frac{m!}{n!}}\e^{-2\left|\alpha\right|^2} \left(2\alpha\right)^{n-m}
  L_m^{n-m}\left(4\left|\alpha\right|^2\right); \: n \geq m, \label {Pi_mn2a} \\
  \Pi_{m,n}\left(\alpha\right) &=\frac{\left(-1\right)^n}{\pi\hbar}
  \sqrt{\frac{n!}{m!}}\e^{-2\left|\alpha\right|^2} \left(2\alpha^*\right)^{m-n}
  L_n^{m-n}\left(4\left|\alpha\right|^2\right); \: m > n. \label {Pi_mn2b}
 \end{align}
\end{subequations}

We are interested in the time evolution of an initial coherent 
state $\ro\left( 0 \right) =\left|\alpha_0\right\rangle \left\langle \alpha_0
\right| $. In this case, the matrix elements of the density operator  
are $\rho_{m,n} = \alpha_0^m\left(\alpha_0^*\right)^n
\e^{-\left|\alpha_0\right|^2}\left(m!n!\right)^{-\frac{1}{2}}$. Substituting this
into (\ref{wigner1}), and reducing the sum in $j$ we have
\begin{eqnarray}
 W\left(\alpha,t\right)&=&\e^{-\left|\alpha_0\right|^2}
 \sum_{l}\sum_{m,n}\frac{\alpha_0^m\left(\alpha_0^*\right)^n}
 {m!n!l!}\sqrt{\left( m+l\right) !\left( n+l\right)!}  \notag \\
 &&\times \gamma_{m-n} ^{l}\left(t\right) \zeta_{m-n} ^{m+n+1}\left(t\right)
 \label{wigner2} \\
 &&\times \exp \left[\left|\alpha_0\right|^2 \gamma_{m-n}\left(t\right)
 +igt\left( m-n\right) \right] \Pi_{m+l,n+l}\left(\alpha\right)\, . \notag
\end{eqnarray}

\section{The classical limit of the diffusive anharmonic oscillator}
\subsection{The equation of motion for the Wigner function and its classical limit}

Taking the Weyl-Wigner transform in both sides of master equation (\ref{master2})
we obtain
a partial differential equation for $W\left(\alpha\right)$. 
The following correspondence formulas \cite{lee} are useful in the
execution of this task:
\begin{eqnarray}
	\a \ro &\rightarrow& \left(\alpha + \frac{1}{2}\delAcc\right) 
	W \left(\alpha \right), \: 
	\ro \a \rightarrow \left(\alpha - \frac{1}{2}\delAcc\right) 
	W \left(\alpha \right),	\label{formww1} \\
	\adag \ro &\rightarrow& \left(\alpha^* - \frac{1}{2}\delA\right) 
	W \left(\alpha \right), \:
	\ro \adag \rightarrow \left(\alpha^* + \frac{1}{2}\delA\right) 
	W \left(\alpha\right),\notag 
\end{eqnarray}
where $\delA \equiv \frac{\partial}{\partial \alpha}$ and 
$\delAcc \equiv \frac{\partial}{\partial \alpha^*}$. The time evolution
of the Wigner function for the diffusive AHO is 
governed by the PDE 
\begin{eqnarray}
	\delt W\left(\alpha,t\right)&=&\left\{-ig\left[\left(2\left|\alpha\right|^2 - 1\right)
	\left(\alpha^* \delAcc - \alpha\delA\right)- \frac{1}{4}
	\left(\alpha^* \delAcc - \alpha\delA\right)\delA\delAcc \right]\notag \right.\\
	&&\left.+ 2\kappa \delA \delAcc \right\} W\left(\alpha,t\right)\label{PDEquartic1}  \: ,
\end{eqnarray}
where $\delt \equiv \frac{\partial}{\partial t} $.

%\subsection{The classical limit}
The classical limit of the AHO can be obtained by taking
the limit $\hbar/J \rightarrow 0$ in Eq. (\ref{PDEquartic1}), 
where $J$ is a characteristic classical action. Assuming that the initial state
is a coherent state with amplitude $\alpha_0$, $J \sim \hbar \left|
\alpha_0\right|^2$. In this case, the classical limit corresponds to take
$\left|\alpha_0\right| \rightarrow \infty$. In order to do this,
let us define a new phase space variable $\beta = \alpha/\left|\alpha_0\right|$.
In terms of this variable, the PDE (\ref{PDEquartic1}) becomes
\begin{eqnarray}
  \delt W\left(\beta, t\right)&=&\left\{-ig\left[\left(2\left|\alpha_0\right|^2
  \left|\beta\right|^2 - 1\right)\left(\beta^* \delBcc - \beta\delB\right)- \frac{1}
  {4\left|\alpha_0\right|^2}
  \left(\beta^* \delBcc - \beta\delB\right)\delB\delBcc \right]\notag \right.\\
  &&\left.+ 2\frac{\kappa}{\left|\alpha_0\right|^2} \delB \delBcc \right\} 
  W\left(\beta,t\right) \: .\label{PDEquartic2}
\end{eqnarray}
Here, $\delB \equiv \frac{\partial}{\partial \beta}$ and 
$\delBcc \equiv \frac{\partial}{\partial \beta^*}$.
However, the constants 
$g$ e $\kappa $ are defined in such way that this limit does not make 
sense, since the nonlinear term is proportional to $\left| \alpha_0\right|^2$ and 
the stochastic sector of this equation vanishes. In order to circumvent 
this problem, we redefine them:
\begin{equation}
  \label{constants}
  g^{\prime} = g\left| \alpha_0\right|^2 \qquad \mathrm{and} 
  \qquad \kappa^{\prime} = \kappa/\left|\alpha_0\right|^2 \:.
\end{equation} 
Proceeding in this way, in the
classical limit, the above PDE becomes a Fokker-Planck (FP) equation
for the classical probability distribution $w \left(\beta,t\right) $:
\begin{equation*} 
	\delt w\left(\beta, t\right)=\left\{
	2ig^{\prime} \left|\beta\right|^2 
	\left(\beta\delB -\beta^* \delBcc\right)        
	+ 2\kappa^{\prime} \delB\delBcc \right\}
	w\left(\beta, t\right)  \: ,
\end{equation*} 
Resorting to
the definition of the constants $g^\prime$ e $\kappa^\prime$ in Eq.
(\ref{constants}), we have
\begin{equation}
	\label{FPquartic2}
	\delt w\left(\alpha, t\right)=\left\{ 
	2ig \left|\alpha\right|^2 
	\left(\alpha\delA -\alpha^* \delAcc\right)          
	+ 2\kappa \delA\delAcc \right\}
	w\left(\alpha, t\right)  \: .
\end{equation}	
Note that the above equation contains only partial derivatives in $\alpha$ and $\alpha^*$
of order one or two. The partial derivatives of superior order, responsible 
for the nonlocal character of Eq. (\ref{PDEquartic1}), vanish in this
limit.

At this point, we introduce the Poisson brackets 
\[
\cP{f}{g} = i\hbar^{-1}\left[\left(\delA f\right)\left(\delAcc g \right)
- \left(\delA g\right)\left(\delAcc f \right)\right]\: ,
\]
and the Eq. (\ref{FPquartic2}) can be rewritten as
\begin{equation}
	\delt w\left(\alpha, t\right)=-\hbar g\cP{\left|\alpha\right|^4}{w\left(\alpha, t\right)}
	-2\hbar^2 \kappa\cP{\alpha}{\cP{\alpha^*}{w\left(\alpha, t\right)}}
	\label{FPquartic3}  \: .
\end{equation}

\subsection{The algebraic structure of the time evolution equation for the classical
distribution function} 
Our objective is to find the solution of the Cauchy problem
described by Eq. (\ref{FPquartic2}) and the initial 
condition $w\left(\alpha, 0\right)$. 
It is interesting to note that, whereas the Weyl-Wigner transform
maps operators in $\mathcal{B}\left(\mathcal{H}\right) $ 
into functions in $\mathcal{F}\left(\Omega \right)$, sup-op are mapped 
in differential operators acting on $\mathcal{F}\left(\Omega \right)$. 
The main benefit of this mapping is the preservation of the commutation
relations. The generality of the Lie algebraic techniques allows one to
extend the results obtained for a problem involving a particular realization
of a determined Lie algebra to another realization of the same algebra. However,
it is necessary to remember that in the classical limit, this correspondence
can not be complete. Compare, \textit{e.g.}, the time evolution equation for the 
the Wigner function (\ref{PDEquartic1}) and for the classical distribution function
(\ref{FPquartic2}). In the classical version, the nonlinear hamiltonian term 
does not present derivatives in the phase space coordinates
of order superior to two. 

Let us consider, for example, the
sup-op $\mathcal{M}$, that acts on a
generic operator $\hat{O} $ as follows 
\[
\mathcal{M} \hat{O}= \adag \a \hat{O}  \:.
\]
Taking the Weyl-Wigner transform in both sides of above equation, we obtain
\begin{eqnarray*}
\mathcal{W}\left(\mathcal{M}\hat{O}\right) &=& 
\mathcal{W}\left(\adag \a \hat{O}\right) \\
&=& \left[\left|\alpha\right|^2 - \frac{1}{2}
\left(1+\alpha\delA -\alpha^*\delAcc \right) - \frac{1}{4}\delA\delAcc
\right]O\left(\alpha\right).
\end{eqnarray*}
Proceeding in this ways, we obtain the following relations between the sup-op 
and the differential operators:
\begin{eqnarray}
\mathcal{M}&\rightarrow& \left|\alpha\right|^2 - \frac{1}{2}
\left(1+\alpha\delA -\alpha^*\delAcc \right) - \frac{1}{4}\delA\delAcc
\: , \notag \\
\mathcal{P}&\rightarrow& \left|\alpha\right|^2 - \frac{1}{2}
\left(1-\alpha\delA +\alpha^*\delAcc \right) - \frac{1}{4}\delA\delAcc
\: ,\label{corresp} \\
\mathcal{J}&\rightarrow& \left|\alpha\right|^2 + \frac{1}{2}
\left(1+\alpha\delA +\alpha^*\delAcc \right) + \frac{1}{4}\delA\delAcc \:,
\notag \\
\mathcal{R}&\rightarrow& \left|\alpha\right|^2 - \frac{1}{2}
\left(1+\alpha\delA +\alpha^*\delAcc \right) + \frac{1}{4}\delA\delAcc
\notag \:.
\end{eqnarray}
At this point, it is useful to introduce the following differential operators:
\begin{subequations}
\label{opdif}
\begin{align}
	Y_0 =& \alpha^*\delAcc - \alpha\delA  \: , 
	\label{Y0} \\
	Y_z =& \left|\alpha\right|^2 - \frac{1}{4}\delA\delAcc \: , 
	\label{Y3} \\
	Y_+ =& \left|\alpha\right|^2 - \frac{1}{2}
	\left(1+\alpha\delA +\alpha^*\delAcc \right) + \frac{1}{4}\delA\delAcc \:,
	\label{Yplus} \\
	Y_- =& \left|\alpha\right|^2 + \frac{1}{2}
	\left(1+\alpha\delA +\alpha^*\delAcc \right) + \frac{1}{4}\delA\delAcc \:,
	\label{Yminus}
\end{align}
\end{subequations}
In terms of these operators, the classical time evolution equation for the
diffusive AHO (\ref{FPquartic2}) becomes
\begin{eqnarray}
	\delt w\left(\alpha, t\right)&=&\left[
	-\frac{ig}{2} \left(Y_+ + Y_- + 2Y_z\right)Y_0 \right.
	\notag \\ 
	&&\left.+2\kappa\left(Y_+ + Y_- - 2Y_z\right)\right]
	w\left(\alpha, t\right) \label{FPquartic4} \\
	&\equiv& L_\infty w\left(\alpha, t\right) \notag \: , 
\end{eqnarray}
The formal solution of the Cauchy problem is given by the application
of a Lie exponential on the initial condition $w\left(\alpha,0 \right) $,
\begin{equation}
	w\left(\alpha,t \right) = \exp\left(L_\infty t\right)
	w\left(\alpha,0 \right) .\label{formal4}
\end{equation}

We can express the Lie exponential $\exp\left(L_\infty t\right)$ 
as a product of exponentials of which the action on functions in
$\mathcal{F}\left(\Omega \right)$ is known. For this, 
we use the BCH expansion formulas, in analogous way to the
quantum version of the problem. The first step is to determine
the Lie algebra generated by the commutation between the operators
that appear in Eq. (\ref{FPquartic4}).

The operators defined in Eq. (\ref{opdif}) obey the
commutation relations presented in Table \ref{comutop}.
Comparing Tables \ref{comutop} and \ref{comutsupop}, we easily note that the differential
operators in Eq. (\ref{opdif}) yield another representation of the four-dimensional
algebra $\mathcal{A}_4$. Because of this, we can identify a Lie 
$su\left(1,1\right)$ subalgebra defined by the operators
$\left\{Y_z,Y_+,Y_-\right\} $. 
\begin{table}[tbh]
	\centering
		\begin{tabular}{c|c c c c}
		       &$Y_0 $ &$Y_z $ &$Y_+ $ &$Y_- $  \\ \hline
		$Y_0 $ &0      &0      &0      &0       \\ 
		$Y_z $ &0      &0      &$Y_+ $ &$-Y_- $ \\ 
		$Y_+ $ &0      &$-Y_+ $&0      &$-2Y_z$ \\
		$Y_- $ &0      &$Y_- $ &$2Y_z$ &0      
		\end{tabular}
	\caption{Commutation relations between the operators defined in
	Eq. (\ref{opdif}). The $i,j$ table entry is the result of the
	commutation of the operator in the $i $-th row with the operator in
	the $j $-th column.}
	\label{comutop}
\end{table}

In the master equation (\ref{master2}), the unitary nonlinear term introduces
the sup-op $\mathcal{M}^2 - \mathcal{P}^2 = \left(\mathcal{M} - \mathcal{P}\right)
\left(\mathcal{M} + \mathcal{P}\right)$. The inclusion of this sup-op to the 
set considered in Table \ref{comutsupop} yields an infinite Lie algebra when
we evaluate its commutation relations with another sup-op.
The trick used to find the solution of Eq. (\ref{master2})
consists formally in considering the sup-op $\left(\mathcal{M} - \mathcal{P}\right)$
as a c-number, since it commutates with the rest and the 
functions obtained in the expansion of the correpondent Lie exponential are, in
fact, functions of this operator. We can determine the quantum state in
time $t $, $\ro\left(t\right) $, by expanding the initial state $\ro_0 $ 
in terms of the eigenfunctions of  $\left(\mathcal{M} - \mathcal{P}\right)$. 

In the same way, the nonlinear hamiltonian term in 
Fokker-Planck equation (\ref{FPquartic2}) introduces 
products of differential operators and they also yield an infinite Lie algebra. 
However, these products are in the form $Y_z Y_0 $, $Y_\pm Y_0 $. Since
$Y_0 $ commutates with the rest of the elements defined in Eq. (\ref{opdif}), we 
can employ an analogous trick to that employed in the solution of (\ref{FPquartic2}), 
i.e., we can formally consider $Y_0 $ a c-number and evaluate the 
functions in the correspondent Lie series as functions of this operator. However,
the solution $w\left(\alpha,t\right) $ determined in this way will be 
``useful" if the action of these differential operators on elements in $\mathcal{F}
\left(\Omega\right) $ is known. Instead and analogously to the procedure adopted
in the quantum version of the problem, we prefer to find the eigenfunctions of 
$Y_0 $ and to express the initial state in terms of them. If we know how each
operator in Eq. (\ref{opdif}) acts on these eigenfunctions, the classical 
distribution $w\left(\alpha,t\right) $ can be written as an
expansion in terms of them with time dependent coeficients.

\subsection{Eigenfunctions of $Y_0$ }
We can easily determine the eigenfunctions of $Y_0$, since they are
related with the eigenfunctions of the sup-op $\left(\mathcal{M} - \mathcal{P}\right) $
by the Weyl-Wigner transform.
Remembering, the eigenfunctions of this sup-op are  
$\left\{\left|m\right\rangle \left\langle n\right|
\right\}_{m,n =0,1,\ldots }$, with eigenvalues $m - n$. 
By Eq. (\ref{corresp}), the Weyl-Wigner transform of the sup-op
$\left(\mathcal{M} - \mathcal{P}\right) $ yields the differential operator 
$Y_0 $, i.e., $\mathcal{W}\left[\left(\mathcal{M} - \mathcal{P}\right)\hat{O}\right] 
= Y_0 \mathcal{W}\left(\hat{O}\right)$,
where $\hat{O} $ is a generic operator acting on $\mathcal{H}$.
Making $\hat{O} = \left|m\right\rangle \left\langle n\right|$, we have
\[
	\mathcal{W}\left[\left(\mathcal{M} - \mathcal{P}\right)\left|m\right\rangle 
	\left\langle n\right|\right]=Y_0 \mathcal{W}\left(\left|m\right\rangle 
	\left\langle n\right|\right)=\left(m-n\right)\mathcal{W}\left(\left|m
	\right\rangle \left\langle n\right|\right).
\]
Therefore, the eigenfunctions of $Y_0$ are the Weyl-Wigner transform of
$\left|m\right\rangle \left\langle n\right| $, namely $\Pi_{m,n} $,
with eigenvalues $\left(m-n\right) $. 

Since the rest of the differential operators defined in Eq. (\ref{corresp})
are related with the sup-op defined in Eq. (\ref{supop}), and the action
of the last on elements of set $\left\{\left|m\right\rangle \left\langle n\right|
\right\}_{m,n =0,1,\ldots }$ is known, we directly obtain the corresponding
action of the first on the functions $\Pi_{m,n} $.
In fact, we have
\begin{eqnarray}
Y_0 \Pi_{m,n} &=& \left(m - n\right)\Pi_{m,n}
\: , \notag \\
Y_z \Pi_{m,n} &=& \frac{1}{2}\left(m + n 
+ 1 \right)\Pi_{m,n}
\: ,\label{acao} \\
Y_- \Pi_{m,n} &=& \sqrt{mn} \Pi_{m-1,n-1}\:,
\notag \\
Y_+ \Pi_{m,n} &=& \sqrt{\left(m + 1\right)
\left(n + 1\right)} \Pi_{m+1,n+1} \notag  \: . 
\end{eqnarray}

In the problems that we are interested, we need to compare 
the time evolutions of the Wigner function and of the classical
distribution associated to a given initial 
state\footnote{A classical distribution associated to a state
$\ro$ is a probability function that the marginal
distributions coincide with the corresponding ones produced by the Wigner function 
$W = \left(2\pi\hbar\right)^{-1}\mathcal{W}\left(\ro \right)$.} $\ro_0$. For our purposes,
it is interesting to express such function as an expansion in
eigenstates of $Y_0 $. Using the properties of the Weyl-Wigner
transform $\mathcal{W}$ we have
\[
W_0 = W\left(0\right) = \mathcal{W}\left(\ro_0 \right) = \sum_{m,n} \rho_{m,n}
\mathcal{W}\left(\left|m\right\rangle\left\langle n\right|\right) \: .
\]
Hence,
\begin{equation}
W_0 = \sum_{m,n} \rho_{m,n} \Pi_{m,n}. 
\label{expansion}
\end{equation}

\subsection{The time evolution of the classical distribution function }
Consider the initial state $\ro_0 = \sum_{m,n} \rho_{m,n} \left|m\right\rangle
\left\langle n\right| $. If the corresponding Wigner function $W_0 $, given by
Eq. (\ref{expansion}), represents 
a valid classical distribution function, we can make $w_0 = w\left(0\right) = W_0$.
The solution of Eq. (\ref{FPquartic2}) for this initial condition is 
\begin{eqnarray}
	w\left(t \right) &=& \exp\left(L_\infty t\right)w_0
	\notag \\
	&=& \exp\left[G\left(t\right)Y_+\right]\exp\left[2Y_z \ln G_z\left(t\right)
	\right]\exp\left[G\left(t\right)Y_-\right] \label{sol} ,
\end{eqnarray}
where
\begin{subequations}
	\label{parameters}
	\begin{align}
		G_z\left(t\right) =\frac{4\sqrt{ig\kappa Y_0}}{4\sqrt{ig\kappa Y_0} \cosh
		\left(2t\sqrt{ig\kappa Y_0}\right) + \left(4\kappa + igY_0\right)\sinh
		\left(2t\sqrt{ig\kappa Y_0}\right)} \:, \label{Gz} \\
		G\left(t\right)= \frac{\left(4\kappa - igY_0\right)\sinh
		\left(2t\sqrt{ig\kappa Y_0}\right)}{4\sqrt{ig\kappa Y_0} \cosh
		\left(2t\sqrt{ig\kappa Y_0}\right) + \left(4\kappa + igY_0\right)\sinh
		\left(2t\sqrt{ig\kappa Y_0}\right)} \:. \label{G}
	\end{align}
\end{subequations}
Note that $G\left(t\right) $ and $G_z\left(t\right) $ are functions of operator
$Y_0 $. 

The action of $\exp\left[G\left(t\right)Y_-\right] $ on the initial condition
$w_0 $ yields
\[
	\exp\left[G\left(t\right)Y_-\right] w_0 = 
	\sum_j  \sum_{r,s} \rho_{r+j,s+j}\frac{v_{r-s}^j\left(t\right)}{j!} 
	\sqrt{\frac{\left(r+j\right)!\left(s+j\right)!}{r!s!}}
	\Pi_{r,s} .
\]
Here, $v_{r-s} \left(t\right) $ is obtained from the function 
$G\left(t\right) $ substituting the operator $Y_0 $ by $r-s $
in Eq. (\ref{G}).

Since $\Pi_{m,n} $ is eigenfunction of $Y_z $ with eigenvalue
$\frac{1}{2}\left(m+n+1\right)  $, the action of $\exp
\left[2Y_z \ln G_z\left(t\right)\right] $ on this function produces 
\[
	\exp \left[2Y_z \ln G_z\left(t\right)\right] \Pi_{m,n} =
	G_z^{m + n + 1}\left(t\right)\Pi_{m,n}.
\]
Hence,
\begin{eqnarray*}
	\exp \left[2Y_z \ln G_z\left(t\right)\right]
	\exp\left[G\left(t\right)Y_-\right]w_0 &=&
	\sum_j  \sum_{r,s} \rho_{r+j,s+j} u_{r-s}^{r+s+1}\left(t\right)\\
	&&\times \frac{v_{r-s}^j\left(t\right)}{j!} 
	\sqrt{\frac{\left(r+j\right)!\left(s+j\right)!}{r!s!}} \Pi_{r,s} ,
\end{eqnarray*}
where $u_{r-s}\left(t\right) $ is obtained from $G_z\left(t\right) $
substituting the operator $Y_0 $ by $r -s $ in Eq. (\ref{Gz}).

Finally, the action of the exponencial $\exp\left[G\left(t\right)
Y_-\right] $ on the above result gives
\begin{eqnarray*}
	\e^{G\left(t\right)Y_+}\e^{2Y_z \ln G_z\left(t\right)}
	\e^{G\left(t\right)Y_-} w_0 &=& \sum_{j,l}
	\sum_{r,s} \rho_{r+j,s+j} u_{r-s}^{r+s+1} \left(t\right)
	v_{r-s}^{l+j}\left(t\right)\\
	&&\times \frac{\sqrt{\left(r+j\right)!\left(s+j\right)!\left(r+l\right)!
	\left(s+l\right)!}}{l!j!r!s!} \Pi_{r+l,s+l} .
\end{eqnarray*}
For the initial condition (\ref{expansion}), the Fokker-Planck equation
(\ref{FPquartic2}) for the diffusive AHO has
the following solution
\begin{eqnarray}
	w\left(t\right)&=& \sum_{j,l}
	\sum_{r,s} \rho_{r+j,s+j} u_{r-s}^{r+s+1} \left(t\right)
	v_{r-s}^{l+j}\left(t\right) \label{solution2}\\
	&&\times \frac{\sqrt{\left(r+j\right)!\left(s+j\right)!\left(r+l\right)!
	\left(s+l\right)!}}{l!j!r!s!} \Pi_{r+l,s+l} \notag.
\end{eqnarray}
Compare this solution with the corresponding one obtained for the density operator,
Eq. (\ref{solution1}), or for the Wigner function, Eq. (\ref{wigner2}),
in the quantum mechanical version of the problem.
\textit{Mutatis mutandis}, the form of the solutions is identical. 
Substituting the functions $\gamma $ and
$\zeta $ in Eq. (\ref{wigner2}) by $v $ and $u $, we find the
result given in Eq. (\ref{solution2}).

\section{An example}
\begin{figure}
	\centering
	\includegraphics[scale=0.5]{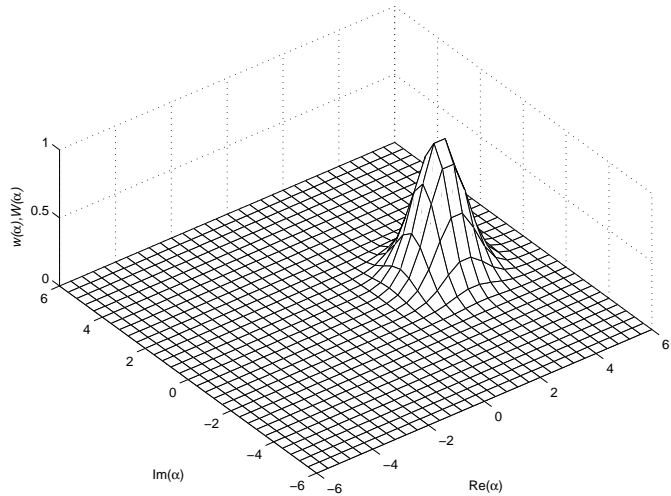}
	\caption{Surface plot of the initial distributions $W\left(\alpha,0\right)$ and
	$w\left(\alpha,0\right)$. In the quantum case, this distribution corresponds to an 
	initial coherent state  $\lvert\alpha_0\rangle\langle \alpha_0\rvert$ 
	with $\alpha_0 = 3$, 
	whereas in the classical case
	it corresponds to a gaussian distribution centered in $\alpha = 3$ with unitary
	variance.}
	\label{fig1}
\end{figure}

In order to gain some insight into the differences between quantum and classical 
dynamics of the nonlinear oscillator, let us compare the
time evolution of a common initial condition. Consider a coherent state
$\lvert \alpha_0\rangle\langle \alpha_0 \rvert$. The associated Wigner 
function in the complex phase space coincides with the corresponding classical
distribution, \textit{i.e.}, a gaussian with variance equal
to the unity centered in $\alpha_0$. An example is shown in Fig. \ref{fig1}. 
The other figures represent the time evolution of this initial state in the
quantum and classical models and for the regimes with and without diffusion.
It is important to mention that the time evolutions were obtained with the
solutions (\ref{wigner2}) and (\ref{solution2}).

\begin{figure}
	\centering
	\includegraphics[scale=0.5]{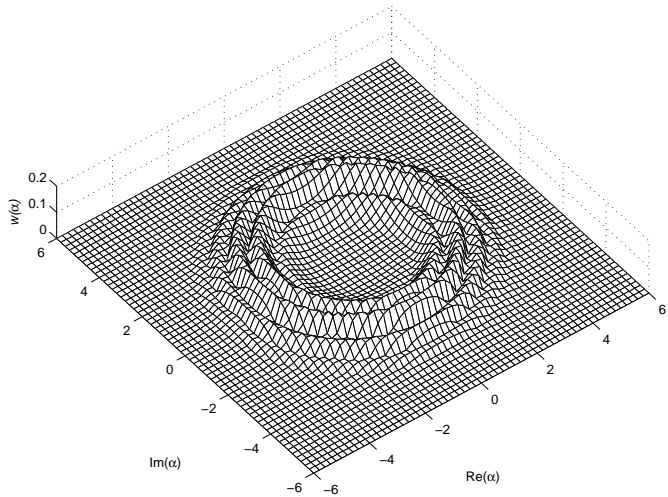}
	\caption{Surface plot of the classical distribution function for the anharmonic oscillator
	at $t=\pi\left(2g\right)^{-1}$ corresponding
	to the hamiltonian evolution ($\kappa= 0$) of the initial condition shown in Fig. \ref{fig1}. 
	For this case, $g/\omega = 0.1$.
	Note the fine-structure yielded by the continuous spiraling of the distribution around
	the origin of the phase space.}
	\label{fig2}
\end{figure}

\begin{figure}
	\centering
	\includegraphics[scale=0.5]{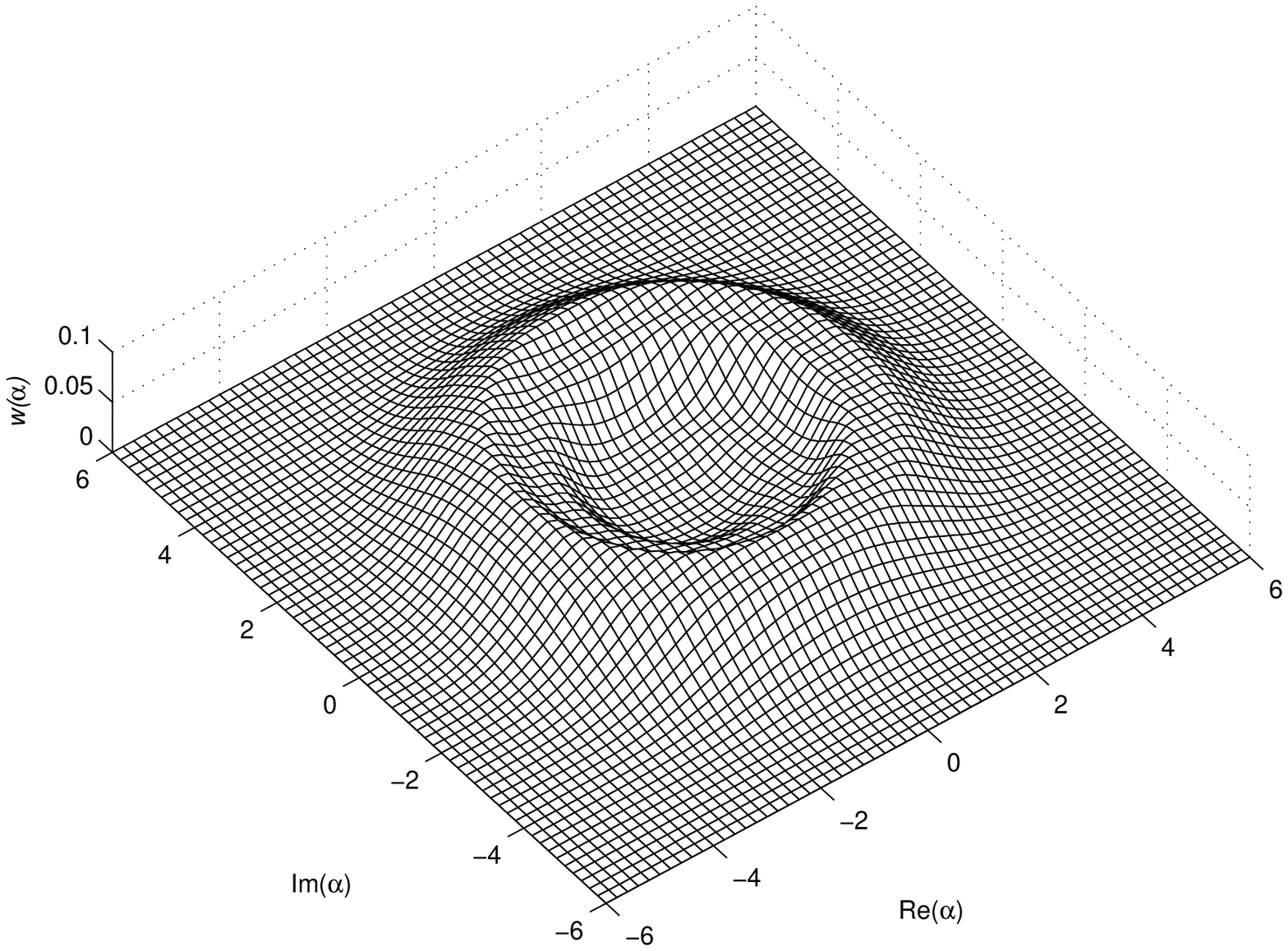}
	\vspace{1cm}
	\includegraphics[scale=0.5]{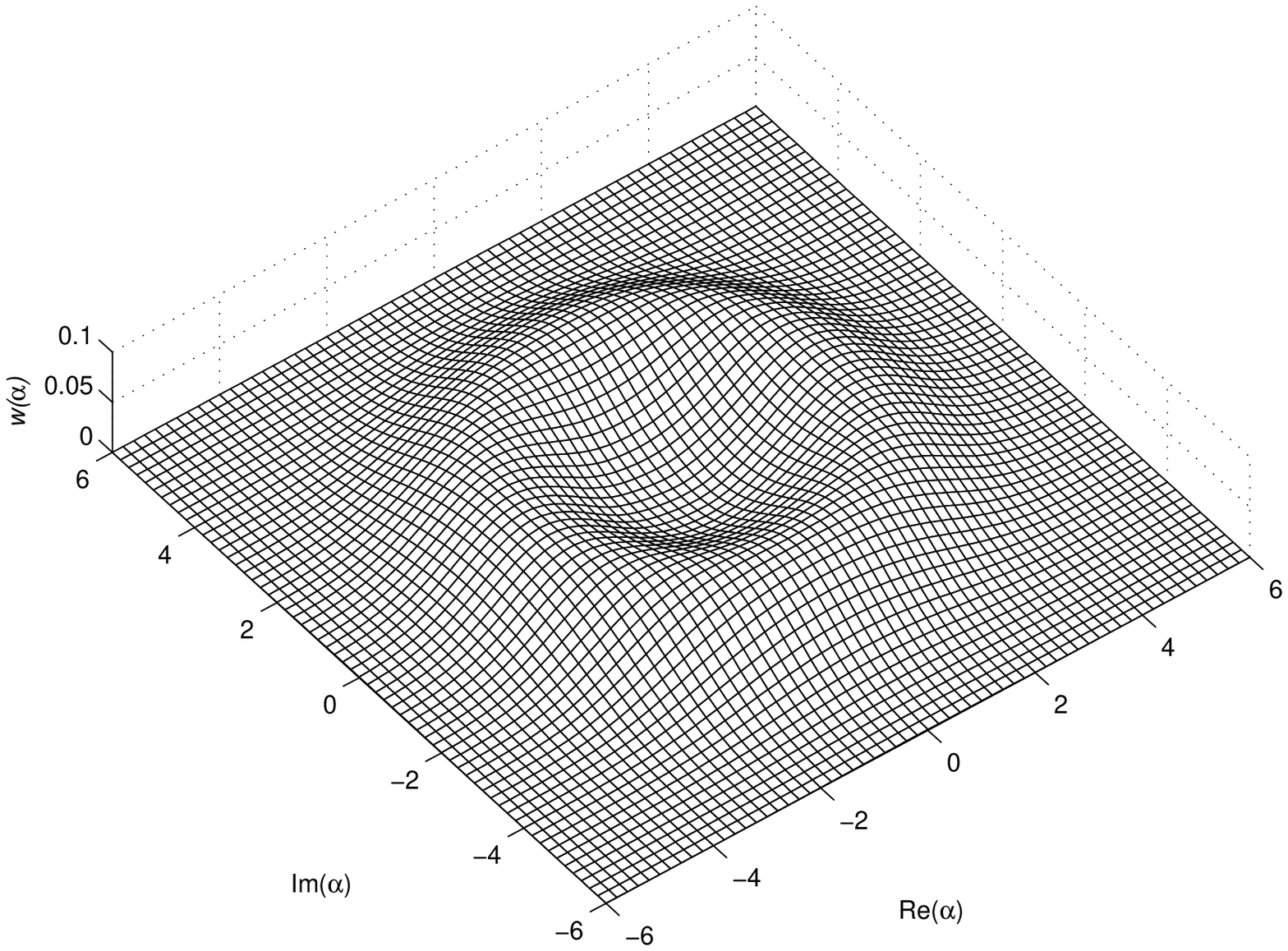}
	\caption{Surface plot of the classical distribution functions for the anharmonic oscillator
	at instants $t=\pi\left(2g\right)^{-1}$ (top) and $t=\pi g^{-1}$ (bottom) 
	corresponding
	to the diffusive evolution ($\kappa/g= 0.1$) of the initial condition shown in Fig. \ref{fig1}. 
	For this case, $g/\omega = 0.1$.
	Note that the fine-structure that should be yielded in the diffusionless regime
	(see Fig. \ref{fig2}) is gradually destroyed.}
	\label{fig3}
\end{figure}

The classical hamiltonian evolution is such that any point of the phase space
moves around the origin with angular frequency proportional to $\left|\alpha\right|^2$,
where $\alpha$ is the coordinate of this point \cite{milburn86}. Therefore, points over the
the initial distribution will rotate with an angular velocity
that depends on their distance to the origin. As consequence, the distribution
will continuously spiral around the origin, as shown in Fig. \ref{fig2}. The
distribution yields a fine-structure in phase space \cite{milburn86}, which is
gradually destroyed if diffusion is included (see Fig. \ref{fig3}). 

\begin{figure}
	\centering
	\includegraphics[scale=0.5]{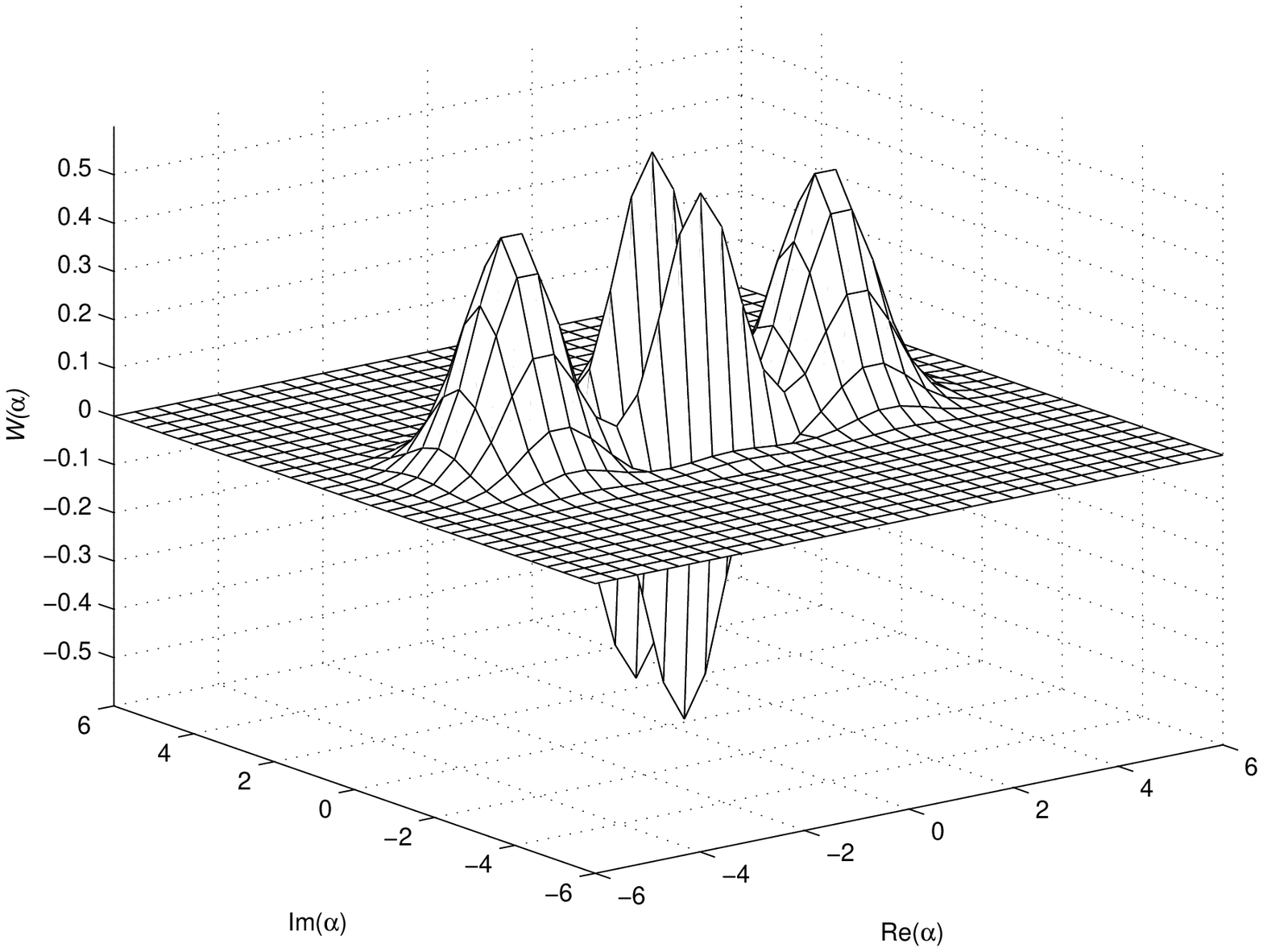}
	\caption{Surface plot of the Wigner function for the quantum anharmonic oscillator
	at $t=\pi\left(2g\right)^{-1}$ 
	corresponding
	to the unitary evolution ($\kappa=0$) of the initial condition shown in Fig. \ref{fig1}. 
	For this case, $g/\omega = 0.1$.
	Note that the appearance of phase space interference due to the coherent superposition
	of states (Schr\"odinger cat state). At this time, the state exhibits 
	some squeezing as well \cite{daniel_milburn}.
	Compare with the corresponding classical one
	shown in Fig. \ref{fig2}.}
	\label{fig4}
\end{figure}

\begin{figure}
	\centering
	\includegraphics[scale=0.5]{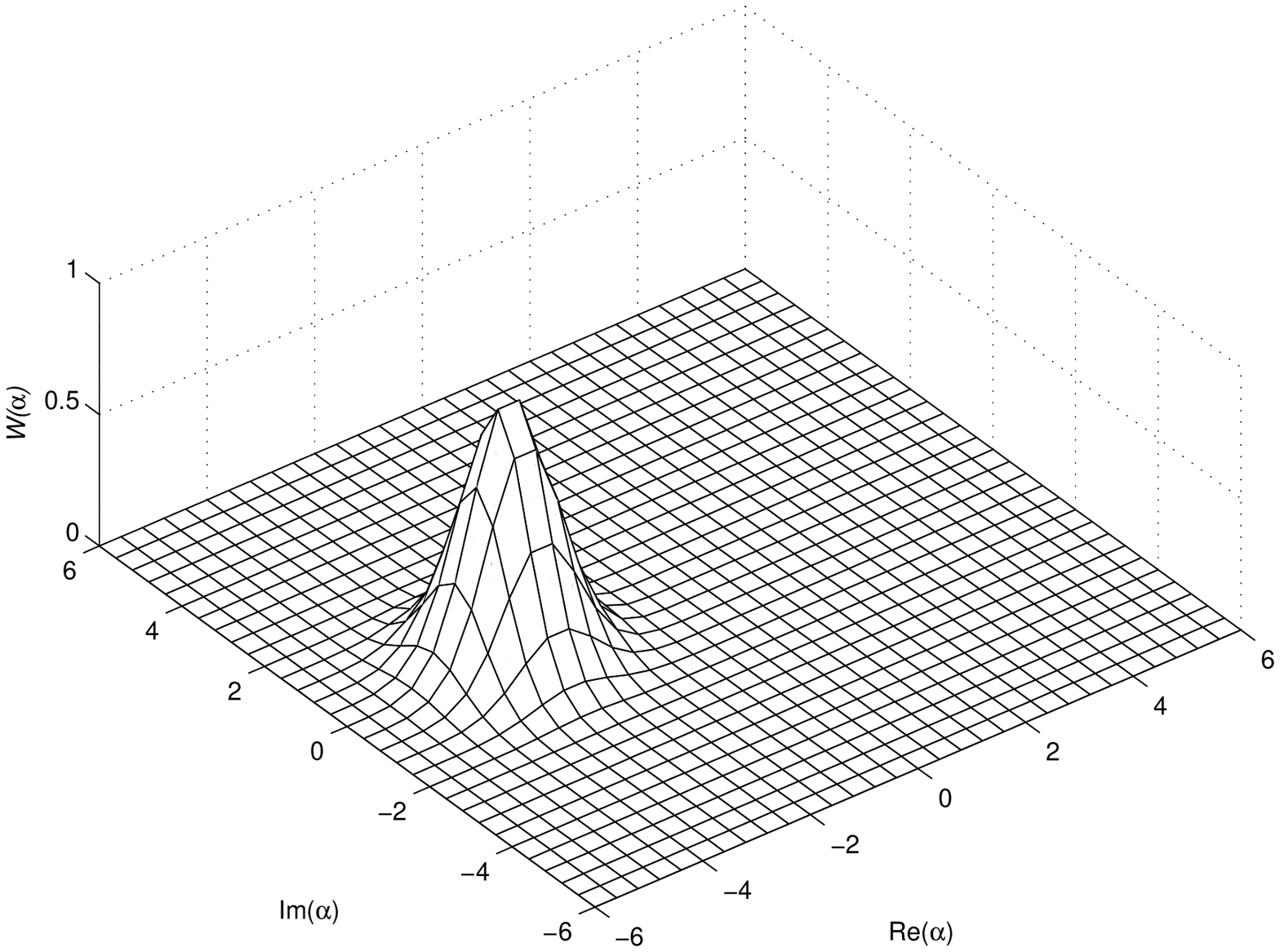}
	\caption{Surface plot of the Wigner function for the quantum anharmonic oscillator
	at $t=\pi g^{-1}$ 
	corresponding
	to the unitary evolution ($\kappa=0$) of the initial condition shown in Fig. \ref{fig1}. 
	For this case, $g/\omega = 0.1$.
	At this instant, the oscillator exhibits an anti-revival since it is found in a 
	coherent state with amplitude
	$-\alpha_0$, where $\alpha_0$ is the amplitude of the initial coherent state.}
	\label{fig5}
\end{figure}

In the quantum version of the model, the unitary evolution of the Wigner
function exhibits a very different behavior that the classical one. For times
equal to $m\pi/\left(2g\right)$, where $m$ is an integer, the nonlinearity leads to 
the quantum superpositions of states ($m$ is odd) or revivals and anti-revivals
($m$ is even). These effects were already reported by Yurke and Stoler 
\cite{yurke86} and examples of them are given in Figs. \ref{fig4} and
\ref{fig5}.

\begin{figure}
	\centering
	\includegraphics[scale=0.5]{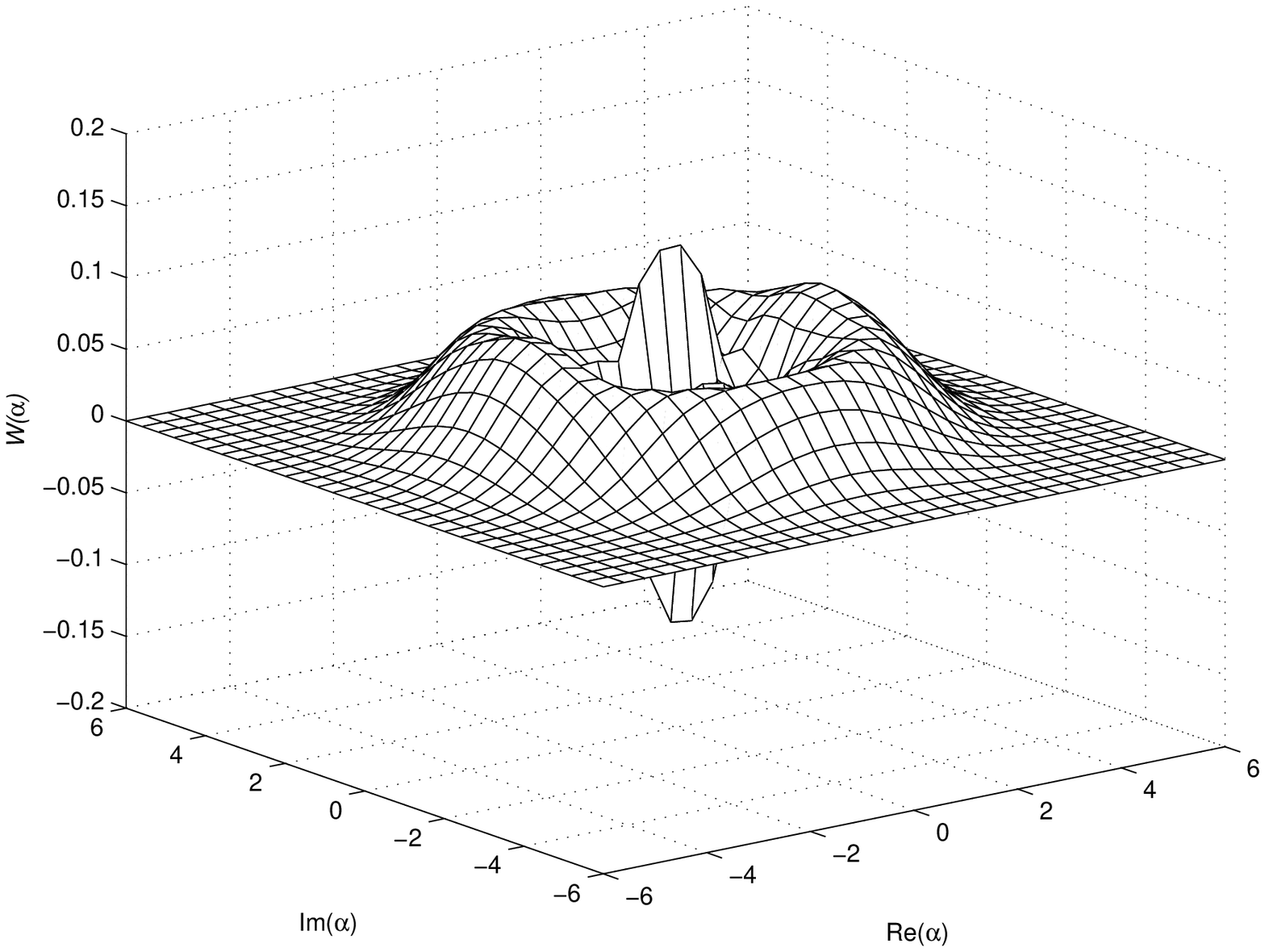}
	\vspace{1cm}
	\includegraphics[scale=0.5]{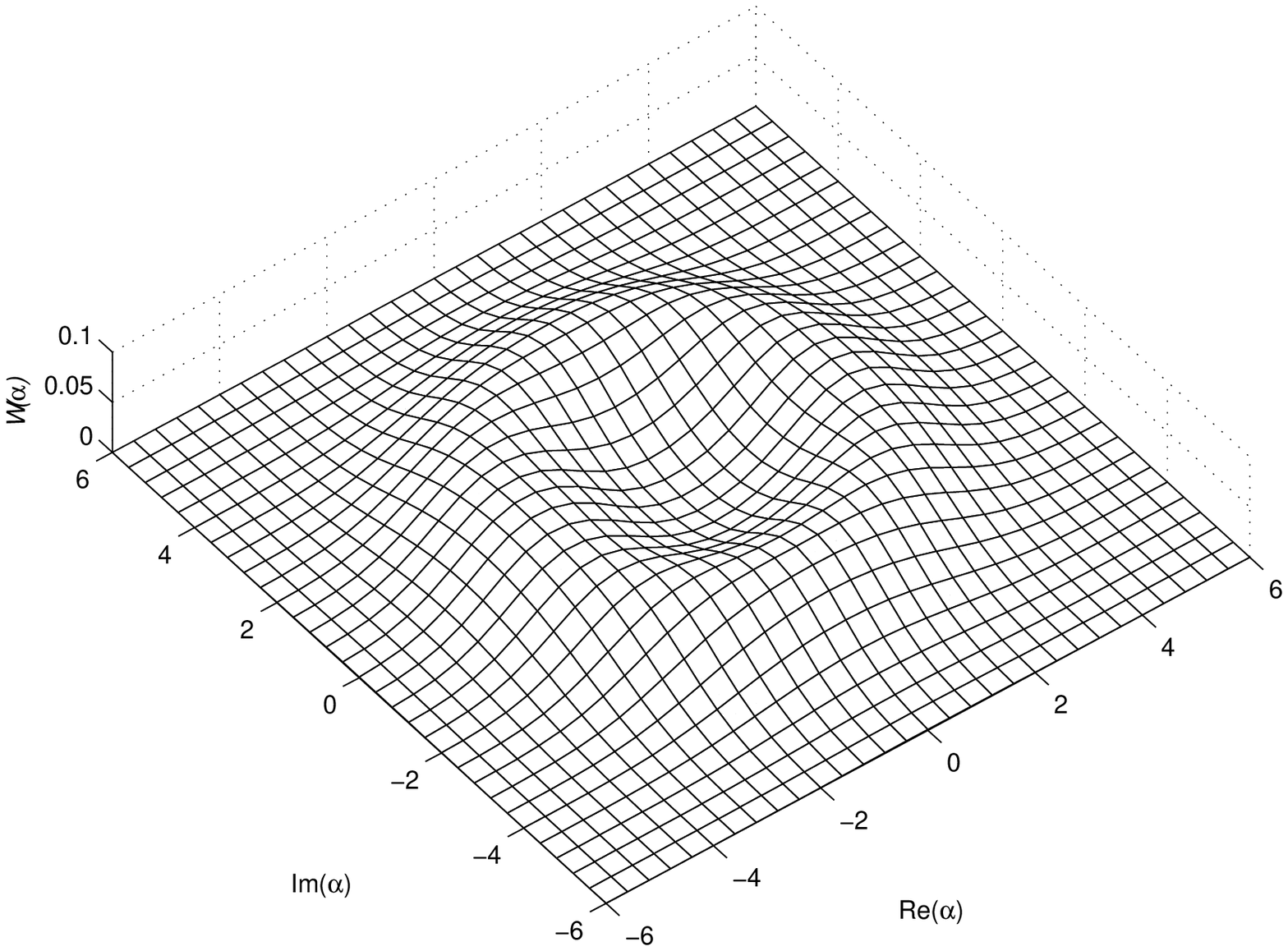}
	\caption{Surface plot of the Wigner function for the quantum anharmonic oscillator
	at instants $t=\pi\left(2g\right)^{-1}$ (top) and $t=\pi g^{-1}$ (bottom)
	corresponding
	to the diffusive evolution ($\kappa/g= 0.1$) of the initial condition shown in Fig. \ref{fig1}. 
	For this case, $g/\omega = 0.1$.
	Comparing with Figs. \ref{fig4} and \ref{fig5} one notes that the
	phase space interferences and the revivals are gradually suppressed.}
	\label{fig6}
\end{figure}

When the diffusion is included, the quantum effects discussed above are gradually
suppressed. The interference in phase space is reduced, and the regions where
the Wigner function is negative diminish (see Fig. \ref{fig6}). For later times,
the Wigner function and the classical distribution take a form of an annular 
volume around the origin of the phase space.
The annular region grows with time but its maximum value diminishes in order to 
mantain constant the integral of $W\left(\alpha,t\right)$ or $w\left(\alpha,t\right)$
over the phase space. This suggests that the Wigner function of the
quantum diffusive AHO converges gradually to the
corresponding classical distribution function.
Non-unitary effects due to quantum dynamics of the open AHO were
investigated by Milburn and Holmes \cite{milburn_holmes}, 
and by Daniel and Milburn \cite{daniel_milburn}, considering the
coupling to a null and non-null temperature reservoir, respectively. In both
works, the authors evaluates the time evolution of the Husimi function ($Q$ function), 
another phase space representation of the density operator. Their results
show a similar behavior for the $Q$ function, in qualitative agreement with
the one reported here.

These results serve to illustrate the procedure and to show that the results are
physically consistent. However, a detailed study of 
the quantum to classical transition in the AHO
must take into account the role played by the parameters of interest,
namely the nonlinearity strengh, the diffusion constant, and
the amplitude of the initial coherent state (a measure of classicallity
of the initial state). This work is in progress.  

\section*{Acknowledgements}
The author thanks to A. C. Oliveira and M. C. Nemes for fruitful discussions 
on this topic.

\end{document}